\documentclass[aps, prl,twocolumn,groupedaddress,showpacs,nofootinbib]{revtex4}
\usepackage{amsmath, amsthm, amssymb}
\usepackage{amscd}
\usepackage{dsfont}
\usepackage{setspace}
\usepackage{graphicx}
\usepackage{color} 
\usepackage{dcolumn}
\begin{document}

\title{Noise-induced looping on the Bloch sphere: Oscillatory effects in dephasing of qubits subject to broad-spectrum noise}
\author{Dong Zhou}
\author{Robert Joynt}
\affiliation{Department of Physics, University of Wisconsin-Madison, Wisconsin
53706, USA}

\begin{abstract}
For many implementations of quantum computing, 1/f and other types of broad-spectrum noise are an important source of decoherence.   An important step forward would be the ability to back out the characteristics of this noise from qubit measurements and to see if it leads to new physical effects.  For certain types of qubits, the working point of the qubit can be varied. Using a new mathematical method that is suited to treat all working points, we present theoretical results that show how this degree of freedom can be used to extract noise parameters and to predict a new effect:  noise-induced looping on the Bloch sphere.  We analyze data on superconducting qubits to show that they are very near the parameter regime where this looping should be observed.
\end{abstract}
\date{\today}

\maketitle

Motivated by the prospect of quantum computation and communication,
coherent quantum operation and control of small systems has become a central
area of physics research.  The isolation of these systems from external
noise is a key problem, as noise produces decoherence.  In solid-state
systems, some level of 1/f or other broad-spectrum noise (BSN) is almost
always present, and is typically difficult to eliminate \cite{kogan}.  
Indeed, in superconducting qubits, {single-electron and other tunneling devices,} this type of noise is recognized as
the factor chiefly responsible for dephasing \cite{wellstood,bialczak,van harlingen, zorin}.

One interesting question is the extent to which the characteristics of the
BSN can be determined by measurements on the qubit itself.  This has been
considered by several authors \cite{schoelkopf, bergli, lutchyn, paladino}.  For the most part, these authors considered
the case of pure dephasing noise. Some theoretical work has been done for "mixed" noise,
which causes both relaxation and dephasing, but this has usually been limited to Gaussian approximations \cite{martinis}, asymptotic analysis, or small numbers of RTNs \cite{galperin}.

In this Letter, we show how to treat “mixed noise” analytically for all times, fully taking into account the non-Gaussian effects.  This will enable us to show how to back out the characteristics of the noise from qubit measurements.  A new physical effect is predicted: noise-induced looping on the Bloch sphere.  We shall analyze data on superconducting flux qubits to show that they are close to the regime in which this effect comes into play.  However, we stress that this effect can occur in any two-level system that is subject to BSN, which can include qubit implementation from atomic and molecular physics as well as solid-state ones.

The effective Hamiltonian of a qubit is often written as $H=-\frac{1}{2}%
\left( \varepsilon \sigma _{z}^{\prime }+\Delta \sigma _{x}^{\prime }\right)
-\frac{1}{2}\vec{h}^{\prime }\left( t\right) \cdot \vec{\sigma}^{\prime }$, 
{ where $\varepsilon$ and $\Delta$ are the energy difference and tunneling splitting between the two physical states.  
For example, in a flux qubit $\varepsilon $ is proportional to the applied flux through the superconducting loop and $\Delta $ is the Josephson coupling.
$\vec{h}^{\prime }\left( t\right)$ is the noise, a random function.} 
  We shall be interested in the case where $\vec{h}^{\prime }\left( t\right) $
comes from $K$ random telegraph noise sources (RTNs) with a wide range of
switching frequencies, giving rise to BSN.  The coordinate system will be
rotated by the angle $\theta=\tan^{-1}(\Delta/\varepsilon)$ so that 
$\sigma_z^\prime=\cos\theta\sigma_z-\sin\theta\sigma_x$ and the qubit energy eigenstates are along the z-axis. $\theta$ is called the working point of the qubit and $H$ is then
\begin{equation}
H=-\frac{1}{2}B_{0}\sigma _{z}-\frac{1}{2}\sum\limits_{k=1}^{K}s_{k}\left(
t\right) \vec{g}_{k}\cdot \vec{\sigma}  \label{eq:hamiltonian}
\end{equation}
where $B_{0}=\sqrt{\varepsilon ^{2}+\Delta ^{2}}.$ $\vec{g}_{k}$ is the coupling
of the $k$-th RTN to the qubit.  $s_{k}\left( t\right) $ switches randomly
between the values -1 and 1 with an average switching rate $\gamma _{k}$. 
Together, these sources generate the random field $\vec{h}\left( t\right)
=\sum_{k=1}^{K}s_{k}\left( t\right) \vec{g}_{k}$.  We assume
throughout that $B_{0}\gg h(t)$.  $ \vec{g}_{k}$ is at an angle $\theta_{k}$ to the
z-axis, which can be thought of as the angle between the noise axis and the
energy axis.  The pure dephasing case, the basis of much work in this
field, corresponds to $\theta _{k}=0$.  
 We deal only
with classical noise, which is generally thought to be appropriate
for the systems of interest here.  Eq.\ref{eq:hamiltonian} is general
enough to describe almost any qubit subject to classical noise, since an
arbitrary power spectrum can be constructed with appropriate choices of
coupling constants $\vec{g}_{k}$ and rates $\gamma_{k}$. 
For simplicity, we drop throughout any azimuthal dependence of the noise and $\vec{g_k}=g_k(\sin\theta_k,0,\cos\theta_k)$.

In many cases, the RTN sources have a fixed direction. 
For example, it has been shown that for superconducting flux qubits the chief
noise source is flux noise \cite{McDermott}, which would put the noise along
the z-axis ($\varepsilon$ direction), and gives $\theta_k =\theta$.
We also stress that the working point $\theta$ is tunable for many different qubit architectures, either by changing $\epsilon$ or $\Delta$ \cite{nakamura, yoshihara,kakuyanagi}.

A convenient solution method for noise from an ensemble of RTNs with an
arbitrary distribution of $\vec{g}_{k}$ and $\gamma _{k}$ has recently been
given \cite{cheng}.  This "quasi-Hamiltonian" method lends
itself to perturbative treatment for both fast (weak-coupling) RTNs: $g_{k}\cos
\theta _{k}\ll \gamma _{k}$ and slow (strong-coupling) RTNs: $g_{k}\cos \theta
_{k}\gg \gamma _{k}$. The occurrence of the trigonometric factor in these
inequalities is crucial: it comes from non-analyticity at the point $%
g_{k}\cos \theta _{k}=\gamma _{k}$ which leads to qualitatively different
behavior in the two regimes.  The method can treat experiments that involve
an arbitrary sequence of control pulses. Because it gives analytic results 
for the non-Gaussian effects of BSN over the whole range of $\theta$ ,
we can identify new systematic effects.

In this work we shall restrict attention to two common experimental
protocols: Energy Relaxation (ER), which measures $n_{ER}\left( t\right)
=\left\langle \sigma _{z}\left( t\right) \right\rangle $, and Spin Echo (SE),
which measures $n_{SE}\left( t\right) =\left\langle \sigma _{x}\left(
t\right) \right\rangle $ with a $\pi$-pulse at $t/2$.  Additional information about the noise
may be available by the use of more complicated pulse sequences \cite{lutchyn}.

 The solutions so obtained are: 
\begin{align}
n_{ER}(t) &\simeq \exp \left[ \left( -2\sum\limits_{m=1}^{M}\gamma
_{m}\epsilon _{2m}^{2}\sin ^{2}\theta _{m}+\Gamma _{1}\right) t\right] 
\label{eq:ER} \\
n_{SE}(t) &\simeq e^{-(\Gamma _{2}+\Gamma _{3})t}\left[ 1+\sum_{m=1}^{M}%
\epsilon _{1m}\sin (g_{m}t\cos \theta _{m})\right] \label{eq:SE}
\end{align}
where $M$ is the total number of slow RTNs and $N$
is the total number of fast RTNs, labeled by $m$ and $%
n$ respectively; $M+N=K$. $\Gamma_{1}$ and $\Gamma _{2}$ are the energy
relaxation and dephasing rates caused by fast RTNs, $\Gamma
_{3}=\sum_{m}\gamma _{m}$ is the overall dephasing rate from slow
RTNs.  $\epsilon _{1m}=\gamma _{m}/\left( g_{m}\cos \theta
_{m}\right) $ and $\epsilon_{2m}=g_m/B_0$
 are the small parameters of the perturbation theory.  For any
value of $\theta$, the portion of the spectrum where the perturbation
theory breaks down is a small fraction of the whole, so we expect the
approximation to be well-controlled.  

The slow RTNs' contribution to the ER signal is much smaller than $
\Gamma _{1}$ and thus can be neglected.  RTN is non-Gaussian noise, which
accounts for the relatively complicated appearance of the formulas.  We
wish to use observations of these signals to back out information on the
distribution of the $\vec{g}$'s and $\gamma $'s.

The fast RTNs' contribution to qubit decoherence is fully captured by
two decay constants $\Gamma _{1}$ and $\Gamma _{2}$.  
For them, our results coincide with Redfield theory \cite{slichter} and 
\begin{align}
\Gamma _{1} &=\sum_{n=1}^N\frac{2\gamma _{n}g_{n}^{2}\sin ^{2}\theta _{n}}{
4\gamma _{n}^{2}+B_{0}^{2}},  \label{eq:Gamma_1} \\
\Gamma _{2} &=\frac{\Gamma _{1}}{2}+\Gamma _{\phi },  \qquad\text{where }
\Gamma _{\phi } =\sum_{n=1}^{N}\frac{g_{n}^{2}\cos ^{2}\theta _{n}}{2\gamma _{n}}.
\end{align}
These fast RTNs give rise to purely exponential decay, as is well
known \cite{astafiev}. 

In contrast, the slow RTNs give rise to the $\Gamma _{3}$ term in
the exponent of $n_{SE}$, which gives exponential decay, but
also to the more complex and $\theta_m$-dependent factors.  These factors
give rise to decays which are quasi-Gaussian over short range of time and possibly oscillations at longer times.

For purposes of fitting, it is necessary for us to specify a not completely
general but yet still flexible model for the noise.  Let $n\left( \gamma
\right) =\sum_{k=1}^{K}\delta \left( \gamma -\gamma_{k}\right) $ be
the distribution of rates and take contant values ${g}_{k}=g$, $\theta_k=\theta$. If there is a range of couplings then $g$ in the following formulas can be regarded as the root-mean-square coupling.  
We will assume a broad noise spectrum by taking $n\left( \gamma \right) =\alpha
\gamma ^{s-1}$ for $\gamma_{\min }<\gamma <\gamma _{\max }$ and $0$ otherwise. Note $s=0$ gives 1/f noise.  
This power-law assumption is often useful for analyzing experimental data while the method itself is capable of treating arbitrary distributions.
The task of data analysis is then to determine $g,\alpha ,s,\gamma _{\min }$, and $\gamma _{\max }$
from observations of $n_{SE}\left( \theta ,t\right) $. 
The data are relatively insensitive to $\gamma _{\max }$, making it difficult to determine.  
We will comment on other limitations of the model below.  The results of the
fitting can be used to make predictions for future experiments.

With these assumptions, the previous results can be simplified into 
\begin{align}
n_{ER}(t)&\simeq e^{-\Gamma _{1}t}, \label{eq:ER_IC}\\
n_{SE}(t)&\simeq e^{-(\Gamma _{2}+\Gamma _{3})t}\left[ 1+\frac{\Gamma _{3}}{
\gamma _{c}}\sin (\gamma _{c}t)\right], \label{eq:Echo_IC}
\end{align}
where $\gamma _{c}=g\cos \theta $ is the critical coupling strength that separates fast and slow RTNs and $\Gamma_3=\int_{\gamma_{min}}^{\gamma_{c}}n(\gamma)\gamma d\gamma$. 
The key point is that by tuning the working point $\theta$, one effectively changes the relative number of fast and slow RTNs in the environment.
Since the two kinds of RTNs have different qualitative effects on the
qubit, we can use this fact to get information about the noise.

Given these results, it is convenient to define 
\begin{align}
\Phi _{SE}\left( t,\theta \right) &=\frac{n_{SE}\left( \theta \right) }{
n_{SE}\left( \theta =\pi /2\right) } \notag\\
&=e^{-\Gamma _{3}t}\left[ 1+\frac{\Gamma _{3}}{\gamma _{c}}\sin \gamma _{c}t
\right],  \label{eq:phise}
\end{align}
where $\Phi _{SE}\left( t,\theta =0\right)$ correspond to the "phase-memory functional" defined by other authors \cite{galperin}.
Note $\Gamma_2(\theta)\simeq \Gamma_1(\theta=\pi/2)/2$, thus it drops out in Eq.\ref{eq:phise}. 

\begin {figure}[tbp]
    \includegraphics*[width=\linewidth]{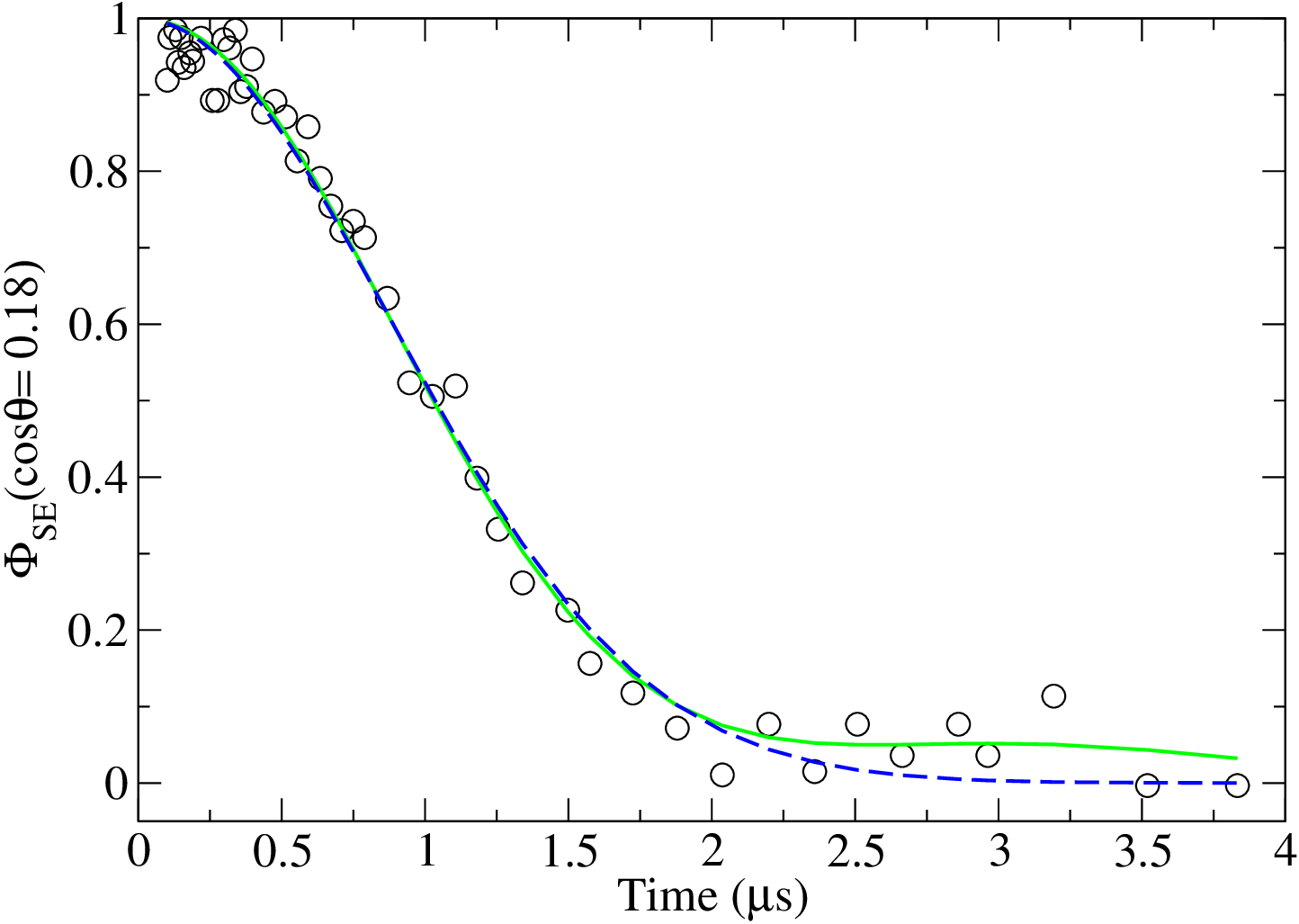}
    \caption{(Color Online) Echo phase memory functional  $\Phi_{SE}$ data in Fig.4A of Ref.\cite{yoshihara}. We fit the $36$ data points to Eq.\ref{eq:phise} (green solid line) and Gaussian model $\Phi_{SE}^G=e^{-\Gamma_G^2t^2}$ (blue dashed line) respectively. The center of the first plateau is thus predicted to be at $\tau_P={5\pi}/{2\gamma_c}\simeq 3.7\mu s$.}\label{fig:echo_4A}
\end{figure}

\begin {figure}[tbp]
    \includegraphics*[width=0.7\linewidth]{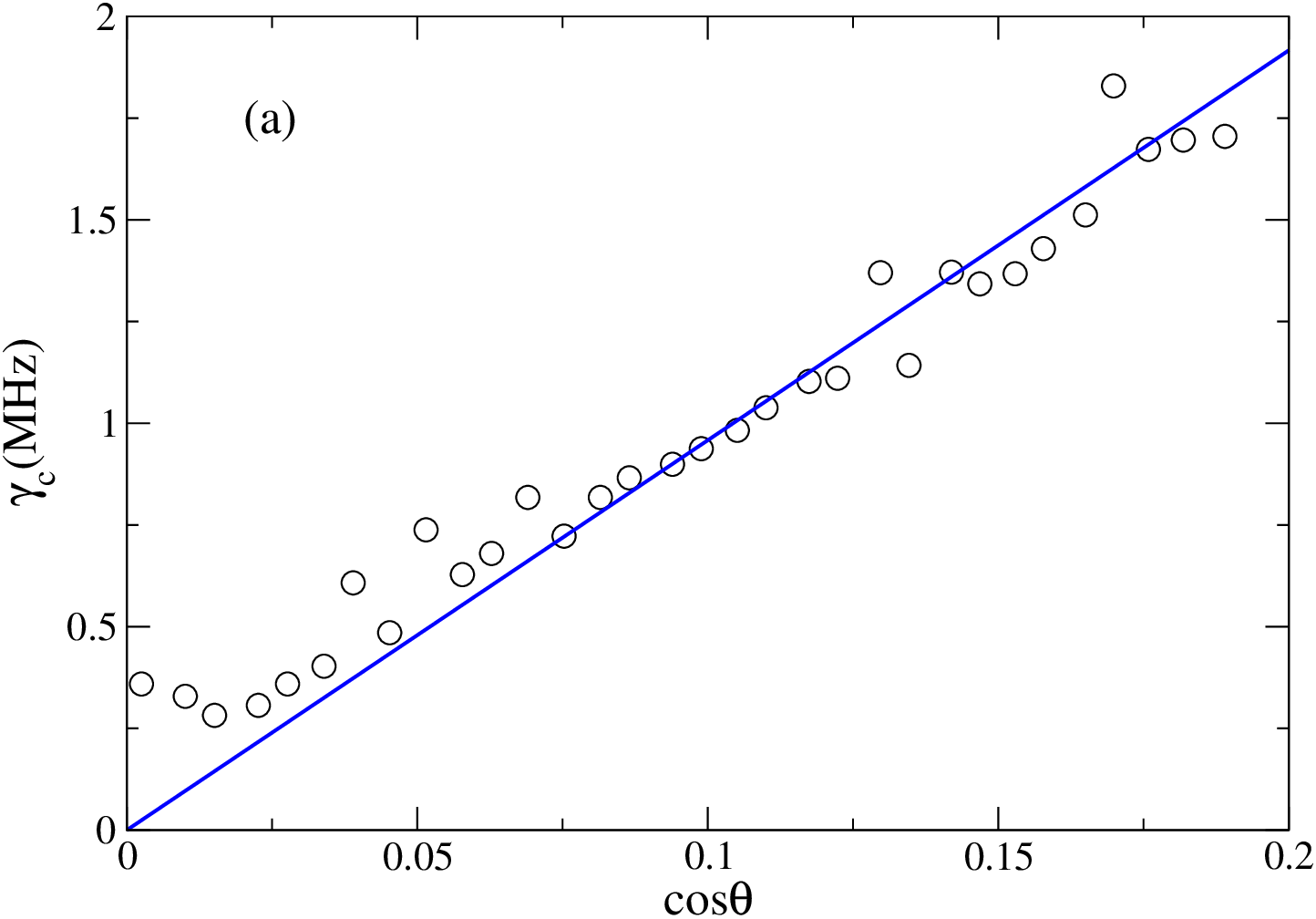}
    \includegraphics*[width=0.7\linewidth]{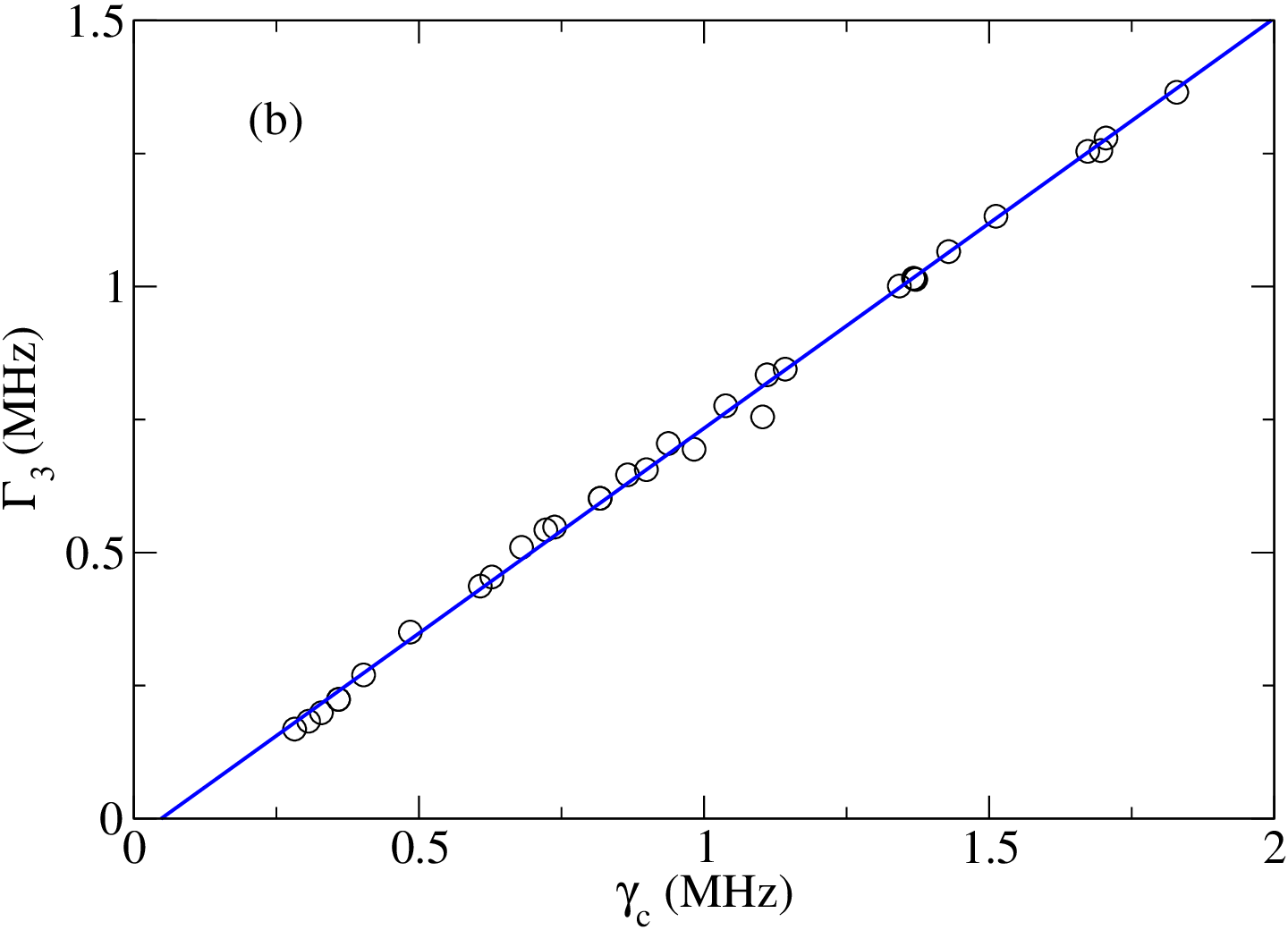}
    \caption{Fitting of sample A data from Ref.\cite{yoshihara}. Both $\gamma_c$ and $\Gamma_3$ are fitted from $\Phi_{SE}$ at various working point. (a) Critical rate $\gamma_c$ versus the working point $\cos\theta$. (b)  Linear regression to $\Gamma_3=\alpha(\gamma_c-\gamma_{min})$.}\label{fig:yoshihara_fit}
\end{figure}

We shall now use these formulas to determine some parameters of the noise
experienced by the flux qubit (sample A) in Ref.\cite{yoshihara}.  For
this system, $\Delta/h =5.445$ GHz and $\varepsilon/h$ varies from $0$ to $1$
GHz, meaning that $\cos \theta $ varies from $0$ to about $0.2$.  Fig.\ref{fig:echo_4A}
shows the fit of Eq.\ref{eq:phise} to the experimental results at $\cos
\theta =0.18$. This gives $\Gamma _{3}=0.99$ MHz and $\gamma_c=2.1$ MHz.  
The fit is certainly excellent, but we note that a Gaussian,
as used by the authors of Ref.\cite{yoshihara} fits equally well over this
time range, with only a hint of deviation at the longest times.  The working point
can be calculated from $\cos\theta=\varepsilon/B_0$, thus fitting $\gamma_c(\theta)$ determines $g/h=9.6$ MHz, as shown in Fig.\ref{fig:yoshihara_fit}(a).
Fitting the data of $\Gamma_3(\theta)$ in Fig.\ref{fig:yoshihara_fit}(b) gives $s=0$ (true 1/f noise), $\alpha =0.77$ and $\gamma _{\min }=0.048$ MHz.  
Note this dependence is very sensitive to the value of $s$ thus can be used to identify the noise distrubition $n(\gamma)$.
We have carried out the analysis for the results of a similar experiment of
Ref.\cite{kakuyanagi}.  All results are given in Table \ref{tab:results}.  An
interesting consistency check is to compute $\Gamma _{1}$ from the model
with the parameters determined since $\Gamma_1$ can be fitted from $n_{ER}(t)$ independently.  For the data of Ref.\cite{yoshihara} (column A of Table \ref{tab:results}), we
find $\Gamma _{1}^{(th)}=0.02$ MHz, while the measured value is $\Gamma
_{1}^{\left( ex\right) }=0.65$ MHz.  This strongly suggests an additional
source of high-frequency noise not captured by our noise model.  For the
data of Ref.\cite{kakuyanagi} (column B), we find $\Gamma _{1}^{(th)}=5.4$ MHz and $%
\Gamma _{1}^{\left( ex\right) }=5.7$ MHz; this shows that the noise model is
reasonably complete for this experiment. 

\begin{table}[b]
\begin{tabular*}{\linewidth}{@{\extracolsep{\fill}}|c|c|c|}
\hline 
 & A & B \tabularnewline
\hline 
$\Delta/h$ (GHz) & $5.445$ & $3.9$\tabularnewline
\hline 
$\varepsilon/h$ (GHz) & $0\sim1$ & $0\sim0.925$\tabularnewline
\hline 
$\Gamma_1^{(ex)}$ (MHz) & $0.65$ & $5.7$\tabularnewline
\hline 
\hline 
$g/h$ (MHz) & $9.6$  & $135$ \tabularnewline
\hline 
$\gamma_{min}$ (MHz) & $0.048$ & $0.098$  \tabularnewline
\hline 
$n(\gamma)$ & $0.77/\gamma$& $0.75/\gamma$\tabularnewline
\hline 
$\Gamma_1^{(th)}$ (MHz) & $0.02$& $5.4$\tabularnewline
\hline 
\end{tabular*}
\caption{Noise characteristics extracted from Ref.\cite{yoshihara} (column A) and Ref.\cite{kakuyanagi} (column B).}\label{tab:results}
\end{table}

The experiments are typically fit by the Gaussian theory for dephasing noise 
\cite{cottet,martinis}.  Galperin \textit{et al}. \cite{galperin} have pointed out
the importance of non-Gaussian effects and have shown numerically that these
effects appear in $\Phi _{SE}\left( t\right) $ at strong coupling and longer
times.  This takes the form of "plateaus".  Eqs.\ref{eq:Echo_IC} and \ref{eq:phise} allow us to quantify and systematically analyze these deviations
from Gaussian behavior at arbitrary $\theta$.

\begin {figure}[tbp]
    \includegraphics*[width=\linewidth]{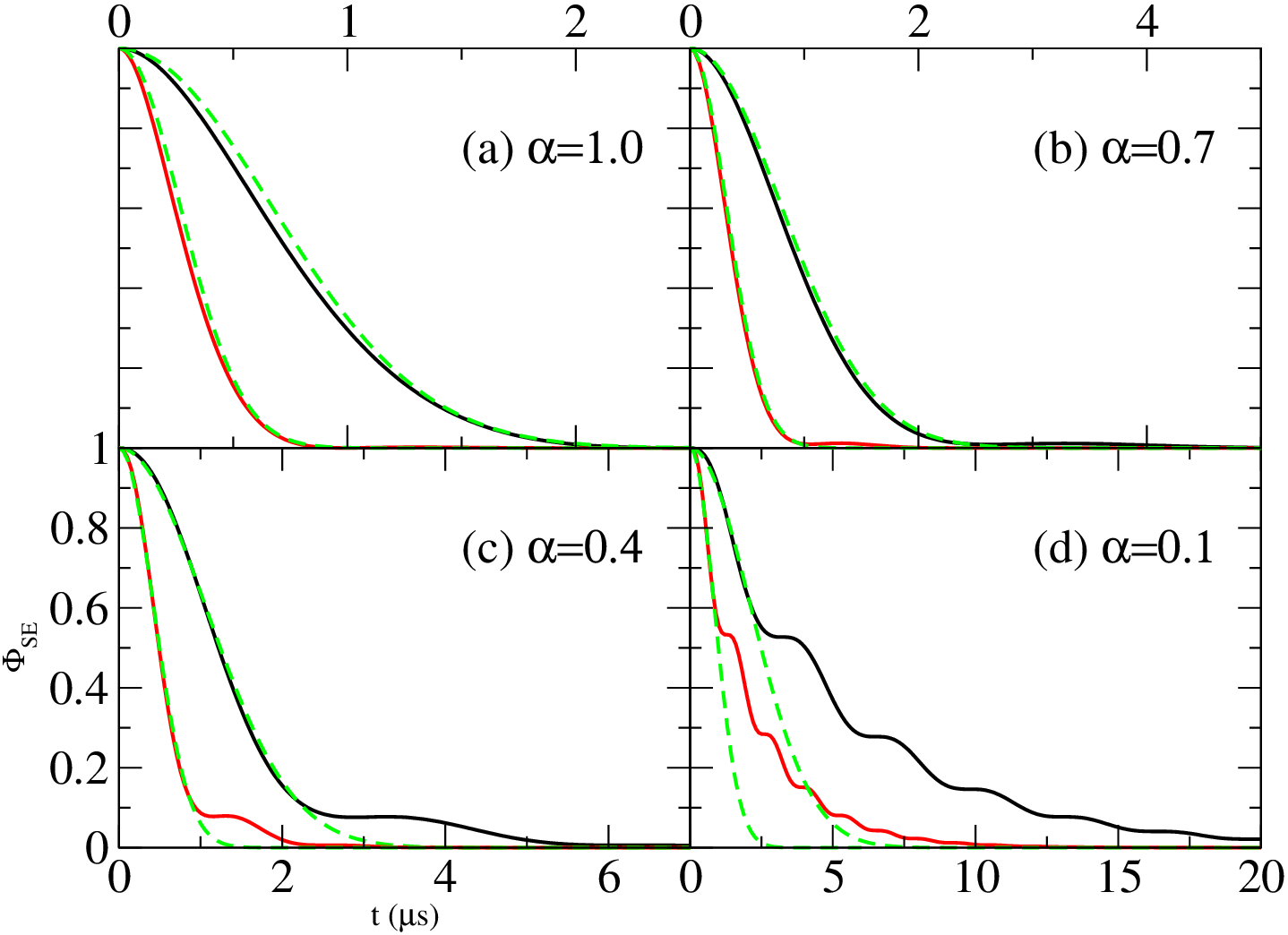}
    \caption{(Color Online) $\Phi_{SE}$ with different noise intensity $\alpha$. In all subfigures, solid black line corresponds to $\cos\theta=0.2$ and solid red line corresponds to $\cos\theta=0.5$.  The corresponding $\Phi_{SE}$ calculated from Gaussian approximation are plotted as dashed green lines. Note  $\Gamma_G\propto \sqrt{\alpha}\cos\theta$ and Gaussian approximation works well in the short time region until the emergence of the first plateau, if there is one. All subfigures share the same axes labels as in (c).}\label{fig:yoshihara_alpha}
\end{figure}

Note that for BSN, $\alpha $ is the parameter that controls the appearance
of the plateaus for 1/f noise. 
Physically, it describes the overall scale of the noise since $K=\int_{\gamma_{min}}^{\gamma_{max}}n(\gamma)d\gamma$.
Fig.\ref{fig:yoshihara_alpha} shows the behavior of $\Phi_{SE}\left( t\right) $ for various values of $\alpha$.  In these plots we
have used the parameters of Ref.\cite{yoshihara} and varied only $\alpha$.
 The experimental value $\alpha=0.77$ is closest to Fig.\ref{fig:yoshihara_alpha}(b).  Fig.\ref{fig:yoshihara_alpha}(c) shows that the experiments are not far from seeing some
non-Gaussian behavior.  

Fig.\ref{fig:yoshihara_alpha} shows that as $\alpha $ is reduced a series of plateaus becomes visible and the phenomenon is better described as oscillations, as would be expected from Eq.\ref{eq:phise}.  
This could be done by reducing the number of RTNs, which is presently a very active area of research \cite{Martinis_2005}.
Fig.\ref{fig:yoshihara_alpha}(d) shows these plateaus quite clearly, and confirms that the period of these oscillations is given by $2\pi/(g\cos\theta)$.
The observation of a period that is adjustable by changing the working point would be an unambiguous signature of the effect. 

The physical interpretation of the oscillations is simple.  $\Phi
_{SE}\left( t\right) $ is a measurement of the probability that the Bloch
vector has returned to its starting point on the equator in the rotating frame.  
The Gaussian theory treats the sphere as a plane and the
decoherence as a random walk on that plane.  In the more rigorous theory,
there is an additional process - a return to the starting point after
looping the sphere. This process is only possible for the slow RTNs 
whose fields can persist over a time long enough for the loop to occur.
It is due to motion along lines of latitude on the sphere and
only the z-component of these fields produces this.  This 
explains why the characteristic looping time is $2\pi/(g\cos\theta)$.

This looping depends on having not too much spread in the distribution of the $\vec{g}
_{k}$.  Clearly a very wide spread would wash out the oscillations, as seen in Eq.\ref{eq:SE}.  On
the other hand, a wide spread that weights the frequency components
of the power spectrum unequally is not consistent with the usual picture
of the origin of 1/f noise as being due essentially to a spread in $\gamma_{k}$.

In conclusion, we have exploited a new mathematical method to obtain the qubit decoherence behavior as a function of working point $\theta$ for qubits that are subject to
BSN coming from classical sources. Varying $\theta$ changes the ratio of strongly and weakly-coupled RTNs. It therefore furnishes a convenient method to back out noise parameters from observations of the qubits themselves. This allows us to refine the picture of non-Gaussian oscillations in qubit decoherence. In particular we have supplied a method to observe the oscillations, and have identified their physical origin as noise-induced looping on the Bloch sphere.  This looping should be observable in many different qubit implementations.

We would like to acknowledge useful discussions with Robert McDermott and Qianghua Wang. Financial support was provided by the National Science Foundation, Grant
Nos. NSF-ECS-0524253 and NSF-FRG-0805045, by the Defense Advanced
Research Projects Agency QuEST program, and by ARO and LPS Grant No. W911NF-08-1-0482.


\begin{thebibliography}{99}
\bibitem{kogan} S. Kogan, \textit{Electronic Noise and Fluctuations in Solids
}, (Cambridge Univ. Press, Cambridge,1996); { M.B. Weissman Rev. Mod. Phys. \textbf{60}, 537 (1988); P. Dutta and P.M. Horn, Rev. Mod. Phys. \textbf{53}, 497 (1981).}
\bibitem{wellstood} F. C. Wellstood, C. Urbina, and J. Clarke, Appl. Phys. Lett. \textbf{99}, 187006 (2007).
\bibitem{bialczak} R. C. Bialczak, R. McDermott, M. Ansmann, M. Hofheinz, N. Katz, E. Lucero, M. Neeley, A.D. O'Connell, H. Wang, A.N. Cleland, and J.M. Martinis, Phys. Rev. Lett. \textbf{99}, 187006 (2007).
\bibitem{van harlingen} D. J. Van Harlingen, T.L. Robertson, B.L.T. Plourde, P.A. Reichardt, T.A. Crane, and J. Clarke, Phys. Rev. B 
\textbf{70}, 064517 (2004).
\bibitem{zorin} {A.B. Zorin, F.J. Ahlers, J. Niemeyer, T. Weimann, H. Wolf, V.A. Krupenin, and S.V. Lotkhov, Phys. Rev. B \textbf{53}, 13682 (1996).} 
\bibitem{schoelkopf} R.J. Schoelkopf, A.A. Clerk, S.M. Girvin, K.W. Lehnert,
and M.H. Devoret, in \textit{Fundamental Problems of Mesoscopic Physics},
ed. Y. Nazarov (Springer, New York, 2004), Ch.1.
\bibitem{bergli}  J. Bergli, Y.M. Galperin, and B.L. Altshuler, Phys. Rev. B \textbf{74}, 024509 (2006).
\bibitem{lutchyn} L. Cywinski, R. M. Lutchyn, C. P. Nave, and S. Das Sarma,
Phys. Rev. B \textbf{77}, 174509 (2008).
\bibitem{paladino} E. Paladino, L. Faoro, G. Falci, and R. Fazio, Phys. Rev.
Lett. \textbf{88}, 228304 (2002);  G. Falci, A. D'Arrigo, A. Mastellone,
and E. Paladino, Phys. Rev. Lett. \textbf{94}, 167002 (2005).
\bibitem{martinis} J. M. Martinis, S. Nam, J. Aumentado, K. M. Lang, and C.
Urbina, Phys. Rev. B \textbf{67}, 094510 (2003).
\bibitem{galperin} Y.M. Galperin, B.L. Altshuler, J. Bergli, and D.V.  Shantsev, Phys. Rev. Lett. \textbf{96}, 097009 (2006); Y.M.  Galperin, B.L. Altshuler, J. Bergli, D. Shantsev and V. Vinokur, Phys. Rev.  B \textbf{76}, 064531 (2007).
\bibitem{McDermott} S. Sendelbach, D. Hover, A. Kittel, M. M\"{u}ck, John M. Martinis, and R. McDermott, Phys. Rev. Lett. \textbf{100}, 227006 (2008).
\bibitem{nakamura} Y. Nakamura, Yu.A. Pashkin, T. Yamamoto, and J.S. Tsai, Phys. Rev. Lett. \textbf{88}, 047901 (2002).
\bibitem{yoshihara} F. Yoshihara, K. Harrabi, A.O. Niskanen, Y. Nakamura,
and J.S. Tsai, Phys. Rev. Lett. \textbf{97}, 167001 (2006).
\bibitem{kakuyanagi} K. Kakuyanagi, T. Meno, S. Saito, H. Nakano, K. Semba,
H. Takayanagi, F. Deppe, and A. Shnirman, Phys. Rev. Lett. \textbf{98},
047004 (2007).
\bibitem{cheng} B. Cheng, Q.-H. Wang, and R. Joynt, Phys. Rev. A \textbf{78}
, 022313 (2008); R. Joynt, D. Zhou, and Q.-H. Wang, arXiv 0906.2843.
\bibitem{slichter} C. P. Slichter, \textit{Principles of Magnetic Resonance}, 3rd ed. (Springer, New York, 1996).
\bibitem{astafiev} O. Astafiev, Y.A. Pashkin, Y. Nakamura, T. Yamamoto, and J.S. Tsai, Phys. Rev. Lett. \textbf{93}, 267007
(2004).
\bibitem{cottet} A. Cottet, Ph. D. thesis (2002).
\bibitem{Martinis_2005} J.M. Martinis \textit{et.al}, Phys. Rev. Lett. \textbf{95}, 210503 (2005).
\end{thebibliography}
 \end{document}